\documentclass[aps,prb,twocolumn,superscriptaddress,amssymb,amsfonts,longbibliography,10pt]{revtex4-2}

\usepackage[latin9]{inputenc}
\usepackage{color}
\usepackage{verbatim}
\usepackage{amsmath}
\usepackage{amssymb}
\usepackage{dsfont}
\usepackage{graphicx}
\usepackage{pbox}
\usepackage{esint}
\usepackage{esvect}
\usepackage{braket}
\usepackage{amsmath,bm}
\usepackage{graphicx}
\usepackage{subfigure}
\usepackage{url}
\usepackage{ulem}
\usepackage{makecell}
\usepackage{lipsum}
\usepackage[dvipsnames]{xcolor}
\usepackage[
    colorlinks,
    linkcolor={blue!80!black},
    citecolor={blue!80!black},
    urlcolor={blue!80!black}
]{hyperref}

\makeatletter
\@ifundefined{textcolor}{}
{%
 \definecolor{BLACK}{gray}{0}
 \definecolor{WHITE}{gray}{1}
 \definecolor{RED}{rgb}{1,0,0}
 \definecolor{GREEN}{rgb}{0,1,0}
 \definecolor{BLUE}{rgb}{0,0,1}
 \definecolor{CYAN}{cmyk}{1,0,0,0}
 \definecolor{MAGENTA}{cmyk}{0,1,0,0}
 \definecolor{YELLOW}{cmyk}{0,0,1,0}
}

\usepackage{pifont}

\newcommand{\hodge}{\,\boldsymbol{\star}\,}
\newcommand{\bk}{\mathbf{k}}

\newcommand{\bn}{\mathbf{n}}
\newcommand{\bb}{\mathbf{b}}
\newcommand{\bJ}{\mathbf{J}}

\newcommand{\bh}{\mathbf{h}}

\newcommand{\bd}{\mathbf{d}}

\newcommand{\cL}{\mathcal{L}}

\newcommand{\bsigma}{\boldsymbol{\sigma}}
\newcommand{\blambda}{\boldsymbol{\lambda}}

\draft
\makeatother
\begin{document}

\title{Orbital magnetization from parallel transport of Bloch states}
\author{Johannes Mitscherling}
\affiliation{Max Planck Institute for the Physics of Complex Systems, N\"othnitzer Str. 38, 01187 Dresden, Germany}
\author{Jan Priessnitz}
\affiliation{Max Planck Institute for the Physics of Complex Systems, N\"othnitzer Str. 38, 01187 Dresden, Germany}
\author{Libor \v Smejkal}
\affiliation{Max Planck Institute for the Physics of Complex Systems, N\"othnitzer Str. 38, 01187 Dresden, Germany}
\affiliation{Max Planck Institute for Chemical Physics of Solids, N\"othnitzer Str. 40, 01187 Dresden, Germany}
\affiliation{Institute of Physics, Czech Academy of Sciences, Cukrovarnick\'a 10, 162 00, Praha 6, Czech Republic}

\date{\today}

\begin{abstract}
Quantum geometric formulations of linear and nonlinear responses can be constructed from a single building block in the form of a gauge-invariant interband transition operator. Here, we identify a second building block for quantum geometry: a band-resolved adiabatic connection operator that captures the noncommutativity between band projectors and their momentum derivatives. The band-resolved adiabatic connection operator, first introduced in the theory of adiabatic driving, serves as a generalized angular momentum within the state manifold of single bands, and we employ it to reformulate expressions for the band-resolved orbital magnetic moment. This form provides a complementary geometric interpretation alongside the multiband separation between energetic- and quantum-state properties by the two-state Berry curvature. Our formalism allows us to present formulas valid for both nondegenerate and degenerate bands, thereby removing the limitations of the common Bloch-state formula. We illustrate our theory by calculating a large orbital magnetization emerging without spin-orbit coupling in a spin-compensated, noncoplanar anomalous Hall magnet with degenerate bands.
\end{abstract}

\maketitle

{\it Introduction.---} Traditional band theory analyzes the properties of metals and semiconductors in terms of band velocities and effective masses. The past few decades have demonstrated that this description is incomplete. Further geometric quantities, such as the quantum metric and Berry curvature, need to be added in order to describe phenomena such as linear and nonlinear Hall and optical responses, superconductivity, flatband, and Landau-level physics~\cite{Nagaosa2010, Xiao2010, Smejkal2022b, Liu2024, Gao2025, Yu2025, Jiang2025, Verma2025}. The geometric quantities capture the momentum-space dependence of wave functions and reveal an intriguing interplay between interband transitions in multiband systems and a single-band description in curved projective space. 

Although the modern theory of orbital magnetization has been revealed more than a decade ago~\cite{Resta2010, Thonhauser2011}, its complete quantum-geometric origin remains elusive. The formulation of the orbital magnetization as a ground state property suggests the need for an additional quantum geometric quantity distinct from the quantum metric and Berry curvature~\cite{Resta2017, Resta2018, Resta2020, Verma2024, Shinada2025}, which depends explicitly on the Bloch Hamiltonian. The orbital magnetization, as a mixed energetic-geometric quantity, is distinct from purely geometric Abelian generalizations, built on traces over band projectors and their derivatives~\cite{Ahn2022, Jankowski2024, Jankowski2024a, Avdoshkin2025, Mitscherling2025, Mehraeen2025}. In addition, the broadly used formulation in terms of Bloch states prevents an efficient investigation of band- and momentum-resolved contributions as well as the treatment of degenerate bands. A gauge-invariant formulation in terms of band projector of individual bands~\cite{Pozo2020, Graf2021, Mera2022, Avdoshkin2023, Antebi2024, Avdoshkin2024a, Avdoshkin2025, Mitscherling2025} will allow the treatment of the orbital magnetic moment for non-degenerate and degenerate bands on equal footing. It could enable band-resolved-geometric insights into the microscopic origin of large orbital magnetization predicted in noncoplanar orbital magnets~\cite{Shindou2001, Hanke2016, Etxebarria2025, Roig2025} and potentially make analytic and numerical investigations~\cite{Lopez2012, Martins2025, Tazuke2025, Cysne2025} more efficient.

\begin{table*}
    \centering
    \begin{tabular}{ccc}
        \hline\hline\\[-2mm]
        & \makecell{Interband \\ transition operator}  & \makecell{Band-resolved adiabatic \\ connection operator} \\
        & \includegraphics[width=4cm]{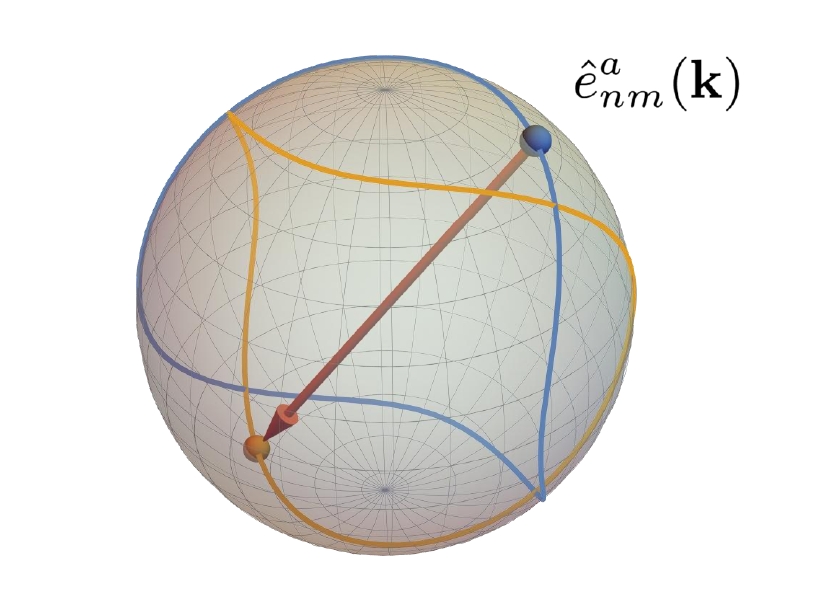} & \includegraphics[width=4cm]{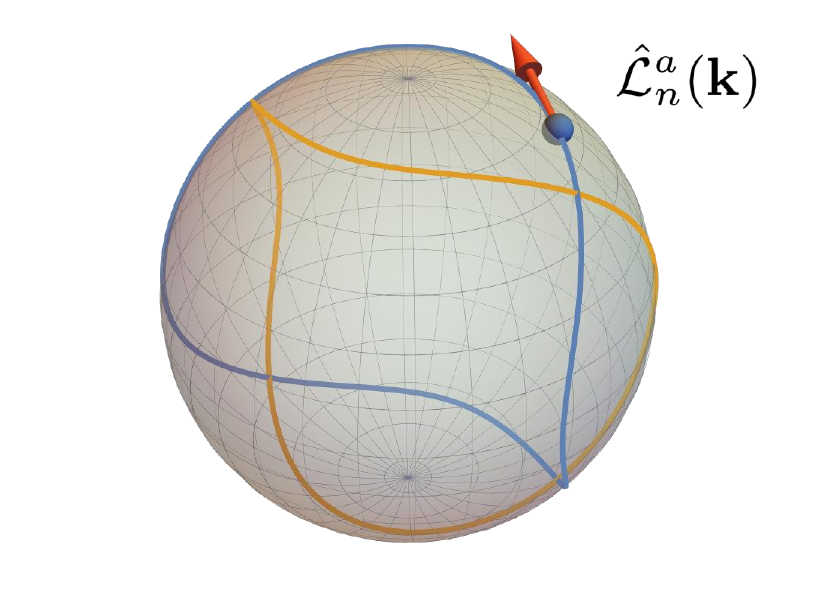} \\
       & \includegraphics[width=4cm]{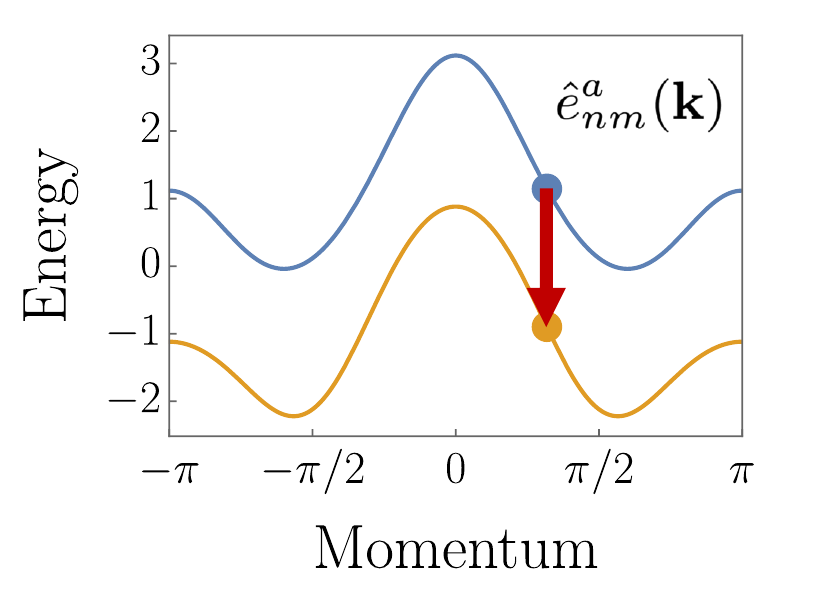}  & \includegraphics[width=4cm]{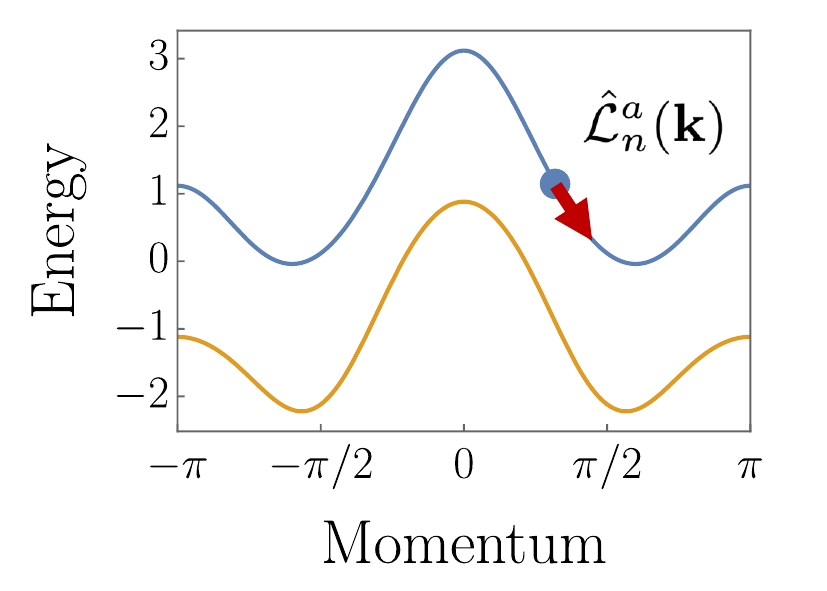} \\[-1mm]\hline\\[-2.5mm]
        Projector form & $i\hat P_n\partial_a\hat P_m\hat P_m$ & $-i\big[\hat P_n,\partial_a\hat P_n\big]$ \\[1mm]\hline\\[-2mm]
        Two-band systems & $\pm\frac{1}{4}\big[\bn\times \partial_a\bn+i\,\partial_a\bn\big]\cdot\bsigma$ & $\frac{1}{2}\big(\bn\times \partial_a\bn\big)\cdot \bsigma$ \\[1mm]\hline\\[-3mm]
        N-band systems & \hspace{0.3cm}\makecell{$-\frac{1}{4}\big(\bb_n\times \partial_a\bb_m\big)\cdot\blambda$ \\[1mm] $+\frac{i}{4}\big(\frac{2}{N}\partial_a\bb_m+\bb_n\hodge \partial_a\bb_m\big)\cdot\blambda$} \hspace{0.3cm} & $\frac{1}{2}\big(\bb_n\times \partial_a\bb_n\big)\cdot \blambda$ \\[3mm]\hline\hline
    \end{tabular}
    \caption{The interband transition operator and band-resolved adiabatic connection operator serve as minimal building blocks for quantum geometry of Bloch states. We provide the gauge-invariant projector form \cite{Ahn2022, Avdoshkin2025, Mitscherling2025} as well as the general expressions for two-band and N-band systems using the generators of SU(2) and SU(N) \cite{Graf2021}. We sketch the physical interpretation within state and energy space for a 1d two-band system. }
    \label{tab:SpinResolvedQGI}
\end{table*}

In this manuscript, we present two complementary quantum-geometric descriptions of the orbital magnetization built upon a systematic quantum geometric decomposition~\cite{Avdoshkin2025, Mitscherling2025, Antebi2024, Ulrich2025} within a fully-quantum finite-temperature Matsubara Green's function approach~\cite{Nourafkan2014}. We show that the orbital magnetization can equivalently be understood as arising from (1) the non-commutativity of parallel Bloch-state transport within the ground state perspective or as being caused by (2) the {\it Abelian} two-state Berry curvature~\cite{Avdoshkin2025, Mitscherling2025} within a multiband perspective. The connection to parallel Bloch-state transport is established upon explicitly summing over virtual band transitions, where we identify a gauge-invariant band-resolved adiabatic connection operator, first introduced in the theory of adiabatic driving~\cite{Kato1950, Nenciu1980, Berry1984, Wilczek1984, Avron1987, Berry2009, Jarzynski2013, Bradlyn2022, Schindler2025}.  The adiabatic connection operator complements the interband transition operator~\cite{Ahn2022, Avdoshkin2025, Mitscherling2025} as a second minimal building block of quantum geometry of Bloch states; and we compare these two building blocks in Tab.~\ref{tab:SpinResolvedQGI}. 

We formulate our theory entirely in terms of band projectors that enable the treatment of non-degenerate and degenerate bands on equal footing. We exemplify theory by investigating the orbital magnetization of a nonrelativistic noncoplanar anomalous (also called topological or geometrical) Hall effect magnet~\cite{Shindou2001, Feng2020, Chen2024} hosting topological two-fold degenerate bands which was studied previously for quantum anomalous~\cite{Zhou2016} and magneto-optical~\cite{Feng2020} effects. Our calculations also confirm that the symmetry of the time-reversal broken anomalous Hall effect and orbital magnetization  are captured by spin group theory~\cite{Watanabe2024} and the effects can emerge without the necessity for spin-orbit coupling and spin-split bands, as pointed out previously~\cite{Feng2020}. Our decomposition of orbital magnetization also shows that the adiabatic connection related contribution can be a dominating contribution in nonrelativistic noncoplanar magnets.  Our results thus demonstrate that nonrelativistic quantum geometry beyond Berry curvature~\cite{Shindou2001}, Berry curvature dipole and quantum metric~\cite{Zhu2025, Ulrich2025} may play an important role in our understanding of observable effects and spintronics applications in noncoplanar antiferromagnets.

{\it Band-resolved adiabatic connection operator.---} We consider a general quadratic Bloch Hamiltonian $\hat H(\bk)$ with arbitary number of bands, which decomposes into the (potentially $N_n$-degenerate) band dispersions $E_n(\bk)$ and the corresponding orthogonal and smooth Bloch projectors $\hat P_n(\bk)$, 
\begin{align}
    \hat H(\bk) = \sum_{n\in \text{bands}} E_n(\bk)\,\hat P_n(\bk) \, .
    \label{eqn:ProjectorExpansion}
\end{align}
The gauge-invariant projectors and their derivatives form the building block of a comprehensive quantum geometric analysis of the Bloch state manifold in terms of physical quantum geometric quantities. The interband transition operator $\hat e^a_{nm}(\bk) = i \hat P_n(\bk)\,\partial_{k_a}\hat P_m(\bk)\,\hat P_m(\bk)$ provides a minimal building block for optical responses~\cite{Ahn2022}, leading to the notion of two-state (or, generally, multi-state) quantum geometry involving band projectors of multiple bands~\cite{Avdoshkin2025, Mitscherling2025}. We point out that the two-state quantum metric and Berry curvature are directly related to the single-band quantities upon summation over all remote bands. In contrast, the interband transition operator is insufficient to provide a complete quantum geometric description of observables that are expected to be ground state properties but yield unconventional virtual interband transitions weighted by energy. We will demonstrate this below explicitly for the orbital magnetic moment. 

For this purpose, we introduce the {\it band-resolved} adiabatic connection operator
\begin{align}
    \hat \cL^a_n(\bk) = -i\big[\hat P_n(\bk),\partial_{k_a}\hat P_n(\bk)\big] \, , \label{eqn:LnDef}
\end{align}
capturing the non-commutativity of the Bloch projectors of a particular band $n$ with its derivative. The adiabatic connection operator is well-established to capture parallel state transport within the state manifold and serves as counterdiabatic Hamiltonian upon summation over all bands in the context of adiabatic quantum driving~\cite{Kato1950, Nenciu1980, Berry1984, Wilczek1984, Avron1987, Berry2009, Jarzynski2013, Bradlyn2022, Schindler2025}. Here, it captures the parallel transport of Bloch states within a given band $n$ under momentum shift, serving as the generator for shifting the band projector $P_n(\bk)$ to $P_n(\bk+\delta\bk)$. As such, the operator $\hat \cL^a_n$ naturally forms an additional minimal building block for physical processes restricted to a band, complementing the interband transition operator $\hat e^a_{nm}$ capturing transitions between two bands; as visualized for a two-band system in Tab.~\ref{tab:SpinResolvedQGI}

It is important to note that both operators are not independent of each other, yielding partially the same information about the Bloch state manifold. The off-diagonal components $\langle u_n(\bk)|\cL^a_n(\bk)|u_m(\bk)\rangle$ for non-degenerate bands yield the non-Abelian Berry connection $r^a_{nm}(\bk) = i\langle u_n(\bk)|\partial_{k_a}u_m(\bk)\rangle$ captured by $\hat e^a_{nm}$ \cite{Ahn2022}. As limiting process of triangulation of the Bloch state manifold \cite{Avdoshkin2023}, the close relation sheds an alternative interpretation on the two-state quantum geometric tensor $\text{tr}\big[\hat e^a_{mn}(\bk)\hat e^b_{nm}(\bk)\big] = (1-\delta_{nm})\,\text{tr}[\hat P_m(\bk)\hat \cL^a_n(\bk)\hat \cL^b_n(\bk)]$ and the quantum geometric tensor $Q^{ab}_n(\bk) = \text{tr}[\hat P_n(\bk)\hat \cL^a_n(\bk)\hat \cL^b_n(\bk)]$. Whereas the former arises from parallel transport of Bloch states with respect to a {\it remote} band, the latter arises from parallel transport of Bloch states with respect to the {\it same} band.  

Considering a generic two-band system $\hat H(\bk) = d_0(\bk) + \bd(\bk)\cdot \bsigma$ expressed in the Pauli basis, the adiabatic connection operator yields the compact form $\frac{1}{2}(\bn(\bk)\times \partial_{k_a}\bn(\bk))\cdot\bsigma$ with $\bn(\bk) = \bd(\bk)/|\bd(\bk)|$. Since $\bn(\bk)$ determines the position on the Bloch sphere and $\partial_{k_a}\bn(\bk)$ is an orthogonal tangent vector to the sphere, $\hat \cL^a_n(\bk)$ generates the rotation on the Bloch sphere, generalizing the notation of the generator of angular momentum to rotations in state space; see Tab.~\ref{tab:SpinResolvedQGI}. Extending this geometric interpretation to general $N$-band Hamiltonians in terms of generalized Bloch vectors $\bb_n(\bk)$~\cite{Graf2021}, we find that $\hat \cL^a_n(\bk) = \frac{1}{2}(\bb_n(\bk)\times \partial_{k_a}\bb_n(\bk))\cdot\blambda$ with the cross product arising from the fully antisymmetric form factors of the SU(N) algebra formed by the generators $\blambda$.

{\it Quantum geometric origin of orbital magnetization.---} We perform a systematic decomposition of the orbital magnetization into its energetic and quantum-geometric components. We start from the expression obtained within the Matsubara Green's function formalism \cite{Nourafkan2014}, which ensures that the orbital magnetization 
\begin{align}
    \mathbf{M}_\text{orb} = -\frac{\partial K}{\partial \mathbf{B}}\Big|_{n,\mathbf{B}=0}
\end{align}
as response of the grand potential per unit volume $K$ under an external magnetic field $\mathbf{B}$ and fixed particle density $n$ yields all virtual interband processes at finite temperature, ensuring the completeness of the quantum geometric decomposition. We note that this approach is valid for non-topological and topological insulators and metals described by the Bloch Hamiltonian $\hat H(\bk)$. 

As described in detail in the SM~\cite{SupplMat}, we insert the projector expansion of the Bloch Hamiltonian in Eq.~\eqref{eqn:ProjectorExpansion} into the Green's function expression isolating the Matsubara summation and the trace over the matrix-valued vertices $\partial_{k_a}\hat H(\bk)$, that decompose into the intraband component $v^a_n(\bk)\hat P_n(\bk)$ and the interband component $i\epsilon_{nm}(\bk)\hat e^a_{nm}(\bk)$ with quasiparticle velocity $v^a_n(\bk) = \partial_{k_a}E_n(\bk)$ and direct band gap $\epsilon_{nm}(\bk) = E_n(\bk)-E_m(\bk)$ \cite{Mitscherling2025}. The intraband contribution vanishes by asymmetry in $x\leftrightarrow y$. The interband contributions result in
\begin{align}
    &M^z = \frac{e}{2\hbar c}\sum_{\underset{n\neq m}{n,m}}\int_\bk f_{nm}(\bk) \big(E_n(\bk)-\mu\big)\,\Omega^{xy}_{mn}(\bk)\nonumber \\&-\frac{e}{2\hbar c}\sum_{\underset{n\neq m}{n,m}}\int_\bk f'_n(\bk)\,\big(E_n(\bk)-\mu\big)\,\epsilon_{nm}(\bk)\,\Omega^{xy}_{mn}(\bk)  \, ,
    \label{eqn:multiBandOMM}
\end{align}
involving the two-state Berry curvature $\Omega_{mn}^{xy}(\bk) = i\,\text{tr}\big[\hat P_n(\bk)\partial_{k_x}\hat P_m(\bk)\partial_{k_y}\hat P_n(\bk)\big]-(x\leftrightarrow y)$, while clearly separating energetic and quantum-geometric contributions. We use the short notation $\int_\bk \equiv \int_\text{BZ}\!\frac{d^d\bk}{(2\pi)^d}$ for the momentum integral over the BZ. The Matsubara summation reveals a second term present at finite temperature \cite{Nourafkan2014}, which is not present in the modern theory of polarization \cite{Resta2010}. In the zero temperature limit, part of Eq.~\eqref{eqn:multiBandOMM} can be brought into $\epsilon_{nm}(\bk)\,\Omega^{xy}_{mn}(\bk)$ summed over occupied bands $n$ and unoccupied bands $m$ \cite{Verma2025}.

From modern theory of orbital magnetization~\cite{Resta2010, Thonhauser2011}, it is well-established that $M^z$ is a ground state property such that the multiband form in Eq.~\eqref{eqn:multiBandOMM} should be reducible to band projectors of the occupied states only, while carefully taking the energetic prefactor into account upon summation. Indeed, the summation over remote bands can be performed in various terms by using $\sum_{m\neq n}\Omega^{xy}_{mn}(\bk) = \Omega^{xy}_n(\bk)$ yielding the Berry curvature of band $n$ \cite{Mitscherling2025}. We find that the remaining terms yield
\begin{align}
    \sum_m E_m(\bk)\,\Omega^{xy}_{mn}(\bk) = i\,\text{tr}\big[\hat H(\bk)[\hat \cL^x_n(\bk),\hat \cL^y_n(\bk)]\big]
    \label{eqn:virtualSummation}
\end{align}
with $\Omega^{xy}_{nn}(\bk) = \Omega^{xy}_n(\bk)$ upon summation over all bands, where we identified the commutator of adiabatic connection operator $\hat \cL^a_n(\bk)$ of band $n$ defined in Eq.~\eqref{eqn:LnDef}. We note that the trace on the right-hand site of Eq.~\eqref{eqn:virtualSummation} is {\it not} a purely {\it Abelian} quantum geometric invariant as it involves $\hat H(\bk)$ within the trace but requires matrix-valued (non-Abelian) quantities. The band-resolved adiabatic connection operator serves as one minimal building block for these. 

\begin{table}
    \centering
    \begin{tabular}{lc}
    \hline\hline\\[-2.5mm]
         & $m^z_n(\bk)\,\big[\frac{e}{\hbar c}\big]$  \\[1mm]\hline\\[-2mm]
        Projector form & \hspace{0.2cm}$-\frac{1}{2}\,\text{Im}\,\text{tr}\Big[\hat H(\bk)\,\big[\partial_{k_x}\hat P_n(\bk),\,\partial_{k_y}\hat P_n(\bk) \big]\Big]$ \hspace{0.2cm}  \\[2mm]\hline\\[-3mm]
         Geometric form & $-\frac{1}{2}\,\text{Im}\,\text{tr}\Big[\hat H(\bk)\,\big[\hat \cL^x_n(\bk),\,\hat \cL^y_n(\bk)\big]\Big]$ \\[2mm]\hline\\[-3mm]
         Bloch state & $\frac{1}{2}\text{Im}\,\langle \partial_{k_x} u_n(\bk)|\hat H(\bk)-E_n(\bk)|\partial_{k_y} u_n(\bk)\rangle$ \\[2mm]
         Two-band & $-\frac{1}{2}|\bd(\bk)|\,\bn(\bk)\cdot\big[\partial_{k_x}\bn(\bk)\times\partial_{k_y} \bn(\bk)\big]$  \\[2mm]
         N-band & $ -\frac{1}{2}\,\bh(\bk)\cdot\big[\partial_{k_x}\bb_n(\bk)\times\partial_{k_y}\bb_n(\bk)\big]$  
    \\[1mm]\hline\hline
    \end{tabular}
    \caption{Different forms of the orbital magnetic moment $m^z_n(\bk)$ of band $n$ in units of $e/\hbar c$ arising from the gauge-invariant projector formalism.}
    \label{tab:summaryOrbitalMagneticMoment}
\end{table}

{\it Orbital magnetic moment.---} By inserting Eq.~\eqref{eqn:virtualSummation} into Eq.~\eqref{eqn:multiBandOMM}, the orbital magnetization restricted to the contributions of each band reads
\begin{align}
    M^z = &\,\,\sum_n\int_\bk \Big[\big(f_n(\bk)+(E_n(\bk)-\mu)f'_n(\bk)\big)\,m^z_n(\bk)\nonumber \\&\hspace{1.5cm}-\frac{e}{\hbar c}f_n(\bk)\,(E_n(\bk)-\mu)\,\Omega^{xy}_n(\bk)\Big] \, .
    \label{eqn:bandresolvedOrbitalMagnetization}
\end{align}
We recover the modern theory of orbital magnetization \cite{Resta2010, Thonhauser2011} at sufficiently low temperature when $(E_n(\bk)-\mu)f'_n(\bk)\approx 0$, after identifying the quantum geometric form of the orbital magnetic moment of each band 
\begin{align}
    m^z_n(\bk)=-\frac{e}{2\hbar c}\,\text{Im}\,\text{tr}\Big[\hat H(\bk)\,\big[\hat \cL^x_n(\bk),\,\hat \cL^y_n(\bk)\big]\Big] \, .
    \label{eqn:orbitalmagneticmomentBand}
\end{align}
From a quantum geometric perspective, the band-resolved orbital magnetic moment yields a clear quantum geometric interpretation as the mismatch of Bloch-state parallel transport in two directions within the curved quantum state manifold of band $n$. 

As summarized in Tab.~\ref{tab:summaryOrbitalMagneticMoment}, the gauge-invariant quantum geometric formulation enables several straightforward reformulations of the orbital magnetic moment~\cite{SupplMat}, which provides further insights into its physical properties and streamlines analytic and numerical evaluations. Using $\hat \cL^x_n(\bk)\,\hat \cL^y_n(\bk) = \partial_{k_x}\hat P_n(\bk)\,\partial_{k_y}\hat P_n(\bk)$ following from basic projector identities, we obtain the projector expression with the minimal number of band projectors. Inserting the band projector of a non-degenerate band $\hat P_n(\bk) = |u_n(\bk)\rangle\langle u_n(\bk)|$ leads to the well-established Bloch state form of the orbital magnetic moment \cite{Resta2010, Thonhauser2011}. For the generic two-band case, we recover the proportionality of the orbital magnetic moment with the Berry curvature, $m_\pm^z(\bk) = \pm \frac{e}{2\hbar c}\Delta(\bk)\Omega^{xy}_\pm(\bk)$ for upper and lower band with band gap $\Delta(\bk)$ \cite{Xiao2007, Yao2008}. As revealed by the general $N$-band form, this proportionality is no longer valid since, in general, the Bloch Hamiltonian vector $\bh(\bk)$ is generically not proportional to the Bloch state vector $\bb_n(\bk)$ \cite{Graf2021}, suggesting a much richer phenomenology of the orbital magnetic moment in systems with more than two bands. 

{\it Orbital magnetization as ground state property.---} We focus on the zero temperature limit in the following, where piecewise smooth $\hat P_\text{occ} = \sum_n \theta(\mu-E_n(\bk))\hat P_n(\bk)$ yields the projector to the occupied states when setting $\theta(0)=1$. The divergence at the Fermi level arising from momentum derivative of $\hat P_\text{occ}(\bk)$ \cite{Resta2017} cancels within the adiabatic connection operator of the occupied states $\hat \cL^a_\text{occ}(\bk) = -i[\hat P_\text{occ}(\bk),\partial_{k_a}\hat P_\text{occ}(\bk)]$ making $\hat \cL^a_\text{occ}(\bk)$ piecewise smooth. We use $\hat \cL^a_\text{occ}$ to define the orbital magnetic momentum of the occupied state manifold analogously to Eq.~\eqref{eqn:orbitalmagneticmomentBand}. We find that the summation over the band-resolved orbital magnetic moments and $m^z_\text{occ}$ deviate by virtual interband transitions within the occupied manifold \cite{SupplMat}
\begin{align}
    \sum_{n\in\text{occ}}m^z_n(\bk) = m^z_\text{occ}(\bk)+\frac{e}{2\hbar c}\!\!\sum_{n,m\in\text{occ}}\!\!\!\!\epsilon_{nm}(\bk) \,\Omega^{xy}_{nm}(\bk) \, .
\end{align}
These transitions are compensated by the deviation between the summation of $E_n(\bk)\Omega^{xy}_n(\bk)$ over occupied states and $\text{tr}[\hat H(\bk)\hat \Omega^{xy}_\text{occ}(\bk)]$ with the matrix-valued (non-Abelian) Berry curvature $\hat \Omega^{xy}_\text{occ}(\bk) = \hat P_\text{occ}(\bk)[\partial_{k_x}\hat P_\text{occ}(\bk),\partial_{k_y}\hat P_\text{occ}(\bk)]\hat P_\text{occ}(\bk)$. Thus, we obtain the compact form of the orbital magnetization in terms of the quantum geometry of the occupied state manifold,
\begin{align}
    M^z &= \int_\bk m^z_\text{occ}(\bk)-\frac{e}{\hbar c} \int_\bk \text{tr}\big[\big(\hat H(\bk)-\mu\big)\,\hat \Omega^{xy}_\text{occ}(\bk)\big] \, .
    \label{eqn:groundstateOrbitalMagnetization}
\end{align}
Upon basic projector manipulations, this expression is equivalent to existing projector expressions of the orbital magnetization~\cite{Ceresoli2006, Lopez2012, Resta2017, Resta2020, Shinada2025}, upon which current ab-initio implementations are based~\cite{Pizzi2020}. These expressions are not minimal in the numbers of projector but enable a more transparent connection to the spatially-local form of orbital magnetization~\cite{Souza2008, Bianco2013}. We recover the Bloch state form of the non-Abelian Berry curvature \cite{Gradhand2011} upon insertion of the respective projector~\cite{SupplMat}.

We close the discussion of the formalism by noting that the smoothness of the Bloch projectors enables a robust approximation of its gradient on a discrete grid $\nabla \hat P(\bk) = \sum_\bb w_b\bb\,\big[\hat P(\bk+\bb)-\hat P(\bk)\big]$ with connecting vectors $\bb$ and adequate weights on the shells $|\bb| = b$ \cite{Marzari1997, Hirschmann2024, Mitscherling2025}, which makes the quantities and its individual terms in Eqs.~\eqref{eqn:multiBandOMM}, \eqref{eqn:orbitalmagneticmomentBand}, and \eqref{eqn:groundstateOrbitalMagnetization} straightforwardly accessible numerically. Furthermore, it enables making connection to the well-established matrix elements $M^{(\bk,\bb)}_{mn} = \langle u_m(\bk)|u_n(\bk+\bb)\rangle$ \cite{Marzari1997} and $C^{(\bk,\bb_1,\bb_2)}_{mn} = \langle u_m(\bk+\bb_1)|\hat H(\bk)|u_n(\bk+\bb_2)\rangle$ \cite{Lopez2012} by inserting the corresponding Bloch projector of the (non-)degenerate band or occupied states.
In the SM \cite{SupplMat}, we provide an estimate for the asymptotic computational complexity for the construction of the Bloch projectors and matrix elements as well as the evaluation of the quantum geometric quantities.

\begin{figure}
    \centering
    \includegraphics[width=1\linewidth]{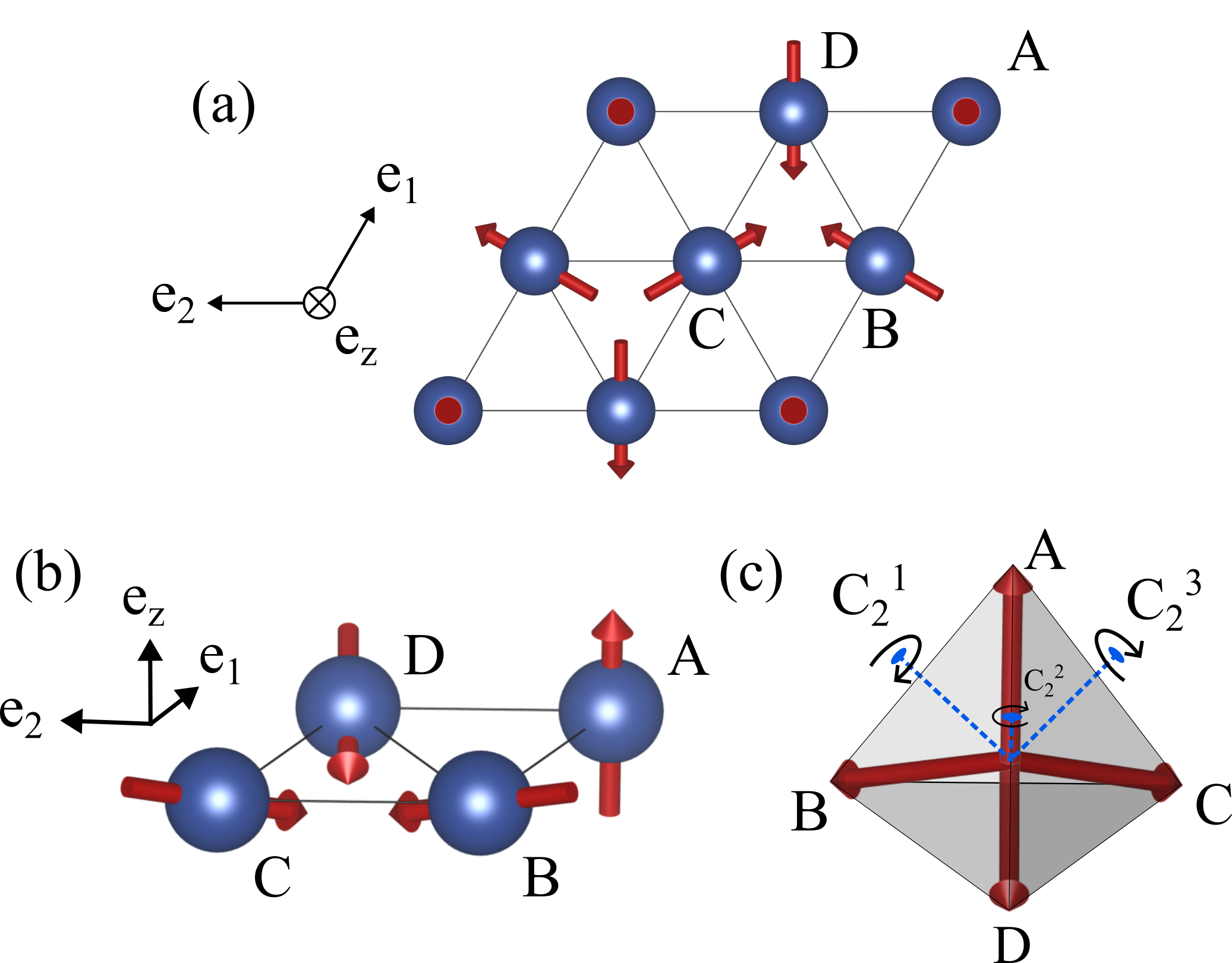}
    \caption{Minimal model of noncoplanar anomalous Hall effect magnet\cite{Feng2020}: magnetic unit cell on a 2D triangular lattice with magnetic moments on 4 sites forming a compensated noncoplanar magnetic order, shown in top-down projection (a) and from a side angle (b). The magnetic moments form a tetrahedron that remains invariant under spin-translation group symmetry operations (two-fold rotations) exchanging pairs of moments (c). }
    \label{fig:NoncollinearMagnetExample}
\end{figure}

\begin{figure*}
    \centering
    \includegraphics[width=0.9\linewidth]{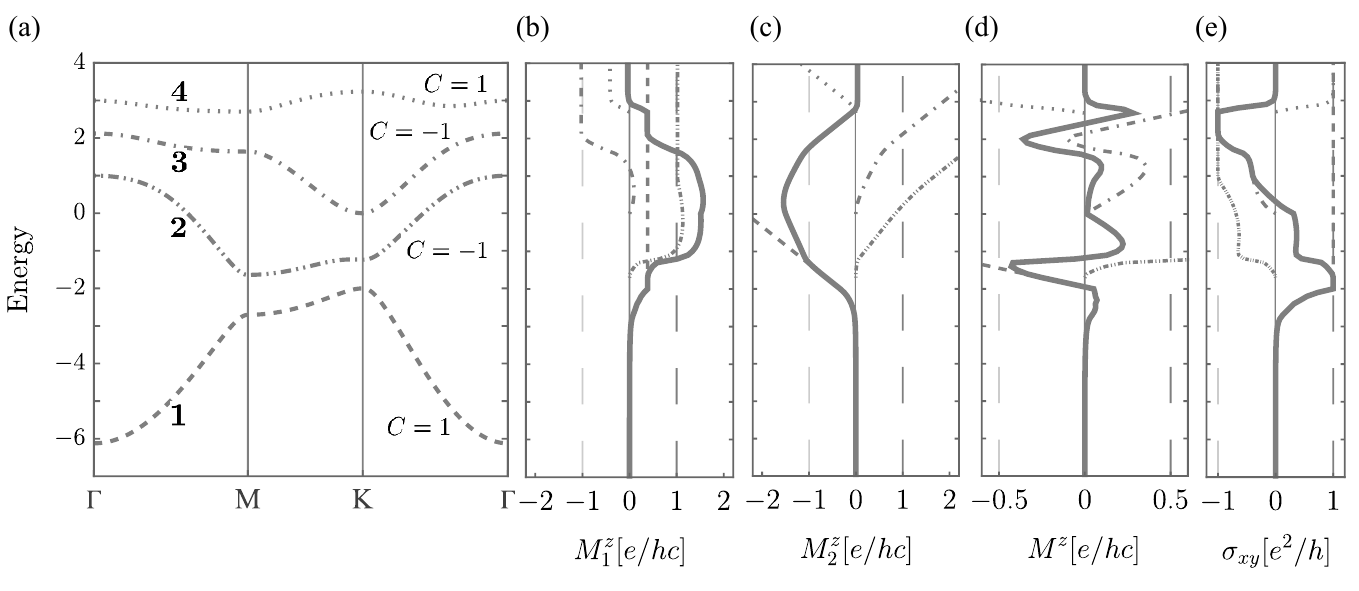}
    \caption{The minimal model hosts four double-degenerate topological bands with nonzero Chern number (a). Both contributions to the orbital magnetization arising from the orbital magnetic moment $m^z_n$ and the Berry curvature $\Omega^{xy}_n$ show large contributions up to $1.5\,e/hc$ of opposite sign (b,c), which results into large and strongly energy-dependent orbital magnetization (d) that strongly deviate from the anomalous Hall response (e)  We give the individual band contributions as dashed and dotted lines.}
    \label{fig:NoncollinearMagnetResults}
\end{figure*}

{\it Application to noncoplanar magnet.---} We illustrate the abstract discussion of our formalism on a minimal model of noncoplanar anomalous (also called topological) Hall effect magnet \cite{Shindou2001,Zhou2016,Feng2020, Smejkal2022b}. For this, we consider a 2D tight-binding model on a triangular lattice~\cite{Feng2020} with nearest-neighbor hopping $t_{ij}$ and on-site exchange coupling $J$,
\begin{align}
    \hat H = -\sum_{\langle ij\rangle,\sigma}\,t_{ij}\hat c^\dagger_{i\sigma}\hat c^{}_{j\sigma} - J\sum_{i,\sigma,\sigma'}\hat c^\dagger_{i\sigma}\bsigma^{}_{\sigma\sigma'}\hat c^{}_{i\sigma'}\cdot\hat \bJ_i
    \label{eqn:modelHamiltonian}
\end{align}
where $\hat c^\dagger_{i\sigma}$ and $\hat c^{}_{i\sigma}$ are the electronic creation and annihilation operators with spin $\sigma = \uparrow,\downarrow$, respectively. The spin quantization axis is chosen in the out-of-plane direction for convenience. The unit cell is composed out of four nonequivalent lattice sites with noncoplanar magnetic moments $\bJ_i=J\hat \bJ_i$ with $|\hat \bJ_i|=1$ that are fully compensated $\sum_i \hat \bJ_i = 0$; see Fig.~\ref{fig:NoncollinearMagnetExample}. We give the $8\times8$ Bloch Hamiltonian matrix in the SM~\cite{SupplMat}. We set the strength of the hopping amplitude and on-site exchange coupling to one in the following, $|t_{ij}|=J=1$.

We analyze the spin symmetries of the model~\cite{Smejkal2022, Watanabe2024, Pengfei2022, Etxebarria2025}. The spin space symmetry group of the model \eqref{eqn:modelHamiltonian} can be written as a product of the spin-only group, spin-translation group and nontrivial spin group $\mathcal{G} = \mathcal{G}_{SO}\times \mathcal{G}_{ST} \times \mathcal{G}_{NT}$. Due to noncoplanar spin ordering, the spin-only group contains only the identity operation $\mathcal{G}_{SO} = \{[E || E]\}$. The nontrivial spin group is $\mathcal{G}_{NT} = P^{3_{001}}6_{001}/^{1}m_{001}^{m_{100}}m_{100}^{m_{100}}m_{120}$, whose orbital part~\cite{Watanabe2024} fully coincides with the crystal space group of the nonmagnetic system $P6/mmm$ (space group no. 191). 
The spin-translation group is a dihedral group isomorphic to the point group $222$~\cite{Feng2020,Watanabe2024} with operations $[C_{2}^1 || E | \tau(\frac{e_1}{2})]$, $[C_{2}^2 || E|\tau(\frac{e_2}{2})]$ and $[C_{2}^3 || E|\tau(\frac{e_1 + e_2}{2})]$ combining two-fold spin-rotation and lattice translations along the hexagonal axes, as indicated in Fig.~\ref{fig:NoncollinearMagnetExample}. 
The spin and orbital part of the spin point group~\cite{Watanabe2024} are $\mathcal{P_\text{s}} = 222 \rtimes 3m = \overline{4}3m$ and $\mathcal{P_{\text{orb}}} = 6/mm'm'$.

In reciprocal space, the spin part $\mathcal{P}_\text{s}$ enforces expectation value of the spin polarization to be zero and bands to be at least doubly degenerate for all momenta, as pointed out previously~\cite{Feng2020,Watanabe2024}. We show the dispersion of the four bands along the high-symmetry lines in Fig.~\ref{fig:NoncollinearMagnetResults}(a), where we label the bands by increasing energy.

According to the nonrelativistic spin group symmetry analysis, both orbital magnetization and anomalous Hall vector~\cite{Smejkal2022b, Etxebarria2025} transform as axial, time-reversal-odd pseudovectors and are determined by the orbital part of the symmetry $\mathcal{P}_\text{orb}$. In our model, non-zero out-of-plane components $\sigma^{xy}$ and $M^z$ are allowed, while in-plane components vanish. Simultaneously, the effective cubic spin symmetry $\mathcal{P}_\text{s}$ enforces spin magnetization to be zero.

\begin{figure*}
    \centering
    \includegraphics[width=0.9\linewidth]{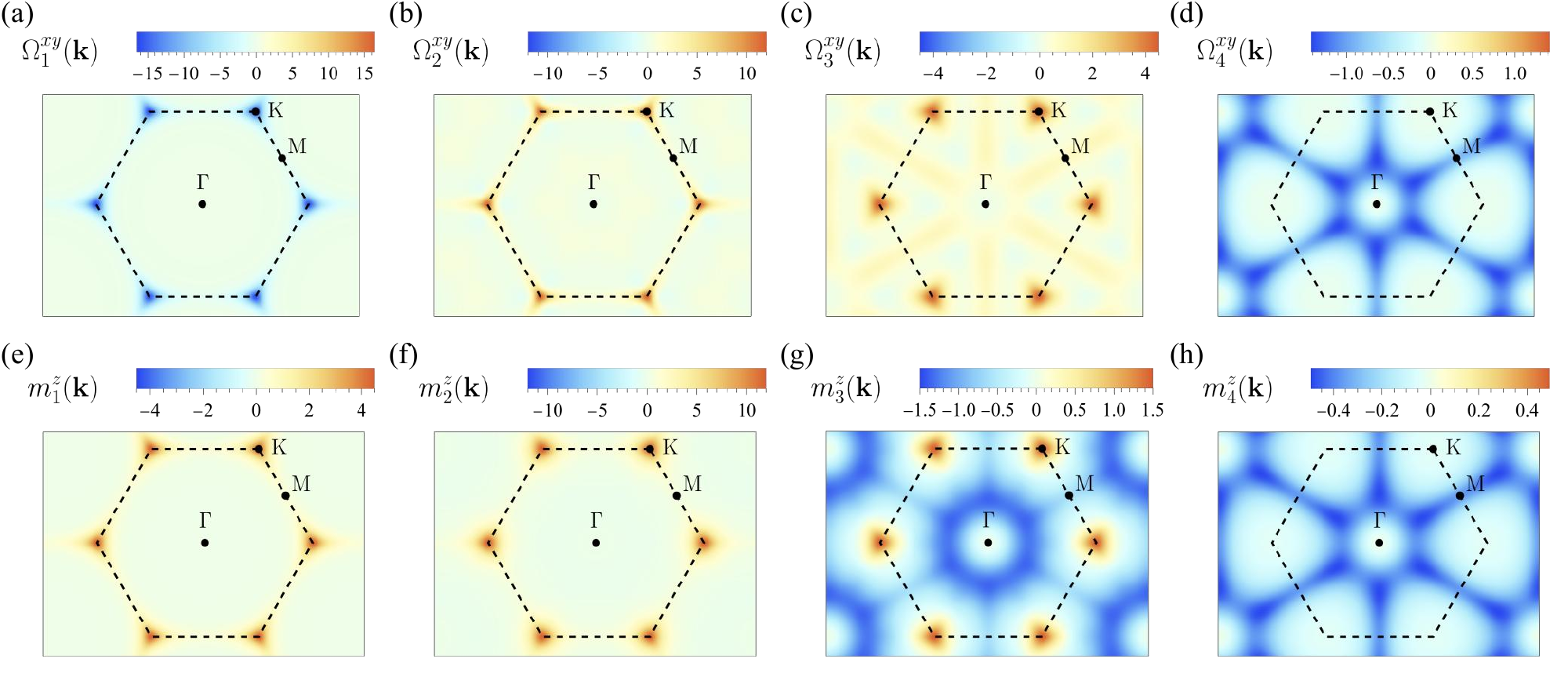}
    \caption{The momentum-resolved Berry curvature $\Omega^{xy}_n(\bk)$ (a-d) and orbital magnetic moment $m^z_n(\bk)$ (e-h) in units $e/\hbar c$ for the four degenerate bands obtained within the projector formalism via the rank-2 projector given in Eq.~\eqref{eqn:projectorDeg}. We denote the BZ and the high-symmetry points and the BZ boundary (dashed lines). The color scale is normalized to the smallest and largest value for each quantity.}
    \label{fig:momentumResolvedFigs}
\end{figure*}

The non-trivial geometry and topology of the degenerate bands are straightforwardly captured within the projector formalism by considering the rank-2 Bloch projectors 
\begin{align}
    \hat P_n(\bk)\! = \!|u_{(n1)}(\bk)\rangle\langle u_{(n1)}(\bk)|\!+\!|u_{(n2)}(\bk)\rangle\langle u_{(n2)}(\bk)| \, ,
    \label{eqn:projectorDeg}
\end{align}
where $|u_{(ns)}(\bk)\rangle$ are orthogonal eigenstates within the energy subspace, i.e., $\hat H(\bk)|u_{(ns)}(\bk)\rangle = E_n(\bk)|u_{(ns)}(\bk)\rangle$. The projector is invariant under SU(2) gauge transformations, unlike the two individual terms. The Berry curvature $\Omega^{xy}_n(\bk) = i\,\text{tr}\big[\hat P_n(\bk)[\partial_{k_x}\hat P_n(\bk),\partial_{k_y}\hat P_n(\bk)]\big]$ as well as the orbital magnetic moment $m^z_n(\bk)$ in Eq.~\eqref{eqn:orbitalmagneticmomentBand} provide the adequate generalization to degenerate bands~\cite{Mera2022}. We note that a naive insertion of $|u_{(ns)}(\bk)\rangle$ into the formulas for non-degenerate bands does, in general, provide non-physical results, with the Abelian Berry curvature as a rare exception. The SU(2) gauge symmetry can be explicitly restored by including further non-Abelian contributions, which are taken into account directly within the projector formalism. 

In Fig.~\ref{fig:momentumResolvedFigs}, we show the momentum distribution of $\Omega^{xy}_n(\bk)$ and $m^z_n(\bk)$ of the four degenerate bands. We note that the three lower dispersive bands yield dominant contributions that arise near the high-symmetry point K, where the three bands are degenerate in the back-folded band structure of the triangular lattice at $J=0$~\cite{SupplMat} and split for finite onsite magnetic moments $J>0$. The large band splitting of the highest band for zero $J$ explains the missing contribution at K. In addition, the upper two bands show contributions that are an order of magnitude smaller but distributed over a broad momentum region on the $\Gamma$-M line, where the upper two bands are degenerate for zero $J$~\cite{SupplMat}. All four bands are topological hosting a nonzero Chern number $C_n = - \frac{1}{2\pi}\int_\text{BZ} d^2\bk\,\Omega^{xy}_n(\bk)$~\cite{Shindou2001, Feng2020}, which explains the fixed relation between large concentrated and small but broadly spread Berry curvature. As both $\Omega^{xy}_n(\bk)$ and $m^z_n(\bk)$ yield the same dependence on momentum changes of the Bloch states via $[\partial_{k_x}\hat P_n(\bk),\partial_{k_y}\hat P_n(\bk)]$, a similar momentum distribution is expected. However, the sign and amplitude can significantly differ, as evident in Fig.~\ref{fig:momentumResolvedFigs}.  

We note that upon (weakly) breaking the symmetry protecting the band degeneracy, the projector of the previously degenerate subspace separate into two well-defined rank-1 projectors, i.e., $\hat P_n(\bk) = \hat P_{(n1)}(\bk)+\hat P_{(n2)}(\bk)$, where the corresponding $\Omega^{xy}_{(ns)}(\bk)$ and $m^z_{(ns)}(\bk)$ can develop a richer substructure and topology, e.g., due to additional avoided band crossings~\cite{Hirschmann2024}. For instance upon adding onsite Zeeman splitting $\hat H_\text{pert} = \delta \sum_i(\hat c^\dagger_{i\uparrow}\hat c^{}_{i\uparrow}-\hat c^\dagger_{i\downarrow}\hat c^{}_{i\downarrow}))$, the four degenerate bands with Chern numbers $\{1,-1,-1,1\}$ split into eight non-degenerate bands with Chern numbers $\{1, 0, -2, 1, 1, -2, 0, 1\}$ for $\delta=0.1$.

We evaluate the two contributions to the orbital magnetization in Eq.~\eqref{eqn:bandresolvedOrbitalMagnetization} in the zero temperature limit and resolved for each degenerate band $M^z_{1,n} = 2\pi\int_\text{occ} m^z_n(\bk)$ and $M^z_{2,n} = -2\pi \int_\text{occ}(E_n(\bk)-\mu)\,\Omega^{xy}_n(\bk)$ in units $e/hc$ integrating over the occupied states $\int_\text{occ} \equiv\int_\bk\,\theta(\mu-E_n(\bk))$; see Fig.~\ref{fig:NoncollinearMagnetResults}(b,c). Summed over all bands, both contributions yield large but non-quantized values of order $1$ for a chemical potential between $-3$ and $3$ dominated by contributions around the high-symmetry point $K$ for the two lower bands. 
The slope of $M^z_{2,n}$ scales as the respective Chern number $C_n$ for fully filled bands according to Eq.~\eqref{eqn:bandresolvedOrbitalMagnetization}~\cite{Resta2010, Thonhauser2011}, whereas the individual band contributions (dashed-dotted lines) of $M^z_{1,n}$ saturate. The opposite relative sign implies a strong dependence on the chemical potential for the full contribution $M^z = M^z_{1}+M^z_{2}$ varying between $-0.4$ and $0.4$ with various sign changes; see Fig.~\ref{fig:NoncollinearMagnetResults}(d). In Fig.~\ref{fig:NoncollinearMagnetResults}(e), we provide band contributions to the anomalous Hall response $\sigma^{xy}_n=-2\pi\int_\text{occ}\Omega^{xy}_n(\bk)$ as well as the full response (solid lines) at zero temperature for varying chemical potential in units $e^2/h$, reflecting the quantization to $\pm 1$ in the band gaps in agreement with the respective Chern numbers~\cite{Feng2020}. It is evident that the anomalous Hall contribution and the orbital magnetization are not simply related confirming previous results for kagome noncoplanar magnets \cite{Hanke2017}.

We point out that the model has been used to explain quantum topological Hall and magneto-optical effect due to scalar spin chirality in $\mathrm{K_{0.5}RhO_2}$~\cite{Zhou2016, Feng2020}. It has been experimentally demonstrated that a similar noncoplanar magnetic structure in $\mathrm{CoTa_3S_6}$ hosts large spontaneous Hall effect~\cite{Takagi2023, Khanh2025}.
We compare spin group symmetries of the model with other noncoplanar magnets which are candidates for studying orbital magnetization from parallel transport from first-principles calculations in the SM~\cite{SupplMat}.

{\it Conclusion and Outlook.---} We have introduced the band-resolved adiabatic connection operator in Eq.~\eqref{eqn:LnDef} into the basic building blocks of quantum geometry for quantum matter and, thereby, provided a direct link to the theory of parallel state transport. We have reformulated the band-resolved orbital magnetic moment and the orbital magnetic moment of occupied states in terms of their respective adiabatic connections, connecting orbital magnetization to the geometry of the curved Bloch-state manifold; see Eq.~\eqref{eqn:orbitalmagneticmomentBand}. We have provided an alternative multiband formulation in terms of the two-state Berry curvature. We have derived equations for the orbital magnetic moment for general two- and N-band systems. Finally, we have shown that the contribution from the adiabatic connection can be large in nonrelativistic anomalous noncoplanar magnets. We have shown that it can originate from contributions broadly distributed over the Brillouin zone away from high-symmetry points, particularly in narrow bands, potentially clarifying the large variety of small or large orbital magnetization seen, for instance, in altermagnetic systems~\cite{Roig2025, Ye2025}. Our formulation might also be advantageous for obtaining exact analytic insights beyond two-band systems, as well as for simplifying the numerical evaluation of band- and momentum-resolved components contributing to the orbital magnetization. We expect our formulation to help gain quantum geometric understanding of other physical properties, e.g., those that are not representable by {\it Abelian} quantum geometric quantities~\cite{Mitscherling2025} only but involve {\it non-Abelian} quantum geometry~\cite{Mera2022, Avdoshkin2024a, Jankowski2024, Liu2025, Chen2025a}.

{\it Acknowledgment.---} We thank Dan S. Borgnia, Marin Bukov, Tobias Holder, Bruno Mera, and Stepan S. Tsirkin for valuable discussions.  J.M. was supported, in part, by the Deutsche Forschungsgemeinschaft under Grant cluster of excellence ct.qmat (EXC 2147, Project No. 390858490). L\v{S} acknowledges funding from the ERC Starting Grant No. 101165122 and Deutsche Forschungsgemeinschaft (DFG) grant no. TRR 288 - 7422213477 (Projects A09 and B05). 

\bibliography{biblio}

@misc{SupplMat,
	title = {{See Supplemental Material (SM) at [URL] for the details on ...  The SM includes Refs..}}}

@article{Graf2021,
  title = {{Berry curvature and quantum metric in $N$-band systems: An eigenprojector approach}},
  author = {Graf, Ansgar and Pi\'echon, Fr\'ed\'eric},
  journal = {Phys. Rev. B},
  volume = {104},
  issue = {8},
  pages = {085114},
  numpages = {19},
  year = {2021},
  month = {Aug},
  publisher = {American Physical Society},
  doi = {10.1103/PhysRevB.104.085114},
  url = {https://link.aps.org/doi/10.1103/PhysRevB.104.085114}
}

@article{Smejkal2022,
  title = {{Emerging Research Landscape of Altermagnetism}},
  author = {\ifmmode \check{S}\else \v{S}\fi{}mejkal, Libor and Sinova, Jairo and Jungwirth, Tomas},
  journal = {Phys. Rev. X},
  volume = {12},
  issue = {4},
  pages = {040501},
  numpages = {27},
  year = {2022},
  month = {Dec},
  publisher = {American Physical Society},
  doi = {10.1103/PhysRevX.12.040501},
  url = {https://link.aps.org/doi/10.1103/PhysRevX.12.040501}
}

@article{Wilczek1984,
  title = {Appearance of Gauge Structure in Simple Dynamical Systems},
  author = {Wilczek, Frank and Zee, A.},
  journal = {Phys. Rev. Lett.},
  volume = {52},
  issue = {24},
  pages = {2111--2114},
  numpages = {0},
  year = {1984},
  month = {Jun},
  publisher = {American Physical Society},
  doi = {10.1103/PhysRevLett.52.2111},
  url = {https://link.aps.org/doi/10.1103/PhysRevLett.52.2111}
}

@article{Gradhand2011,
  title = {Calculating the Berry curvature of Bloch electrons using the KKR method},
  author = {Gradhand, M. and Fedorov, D. V. and Pientka, F. and Zahn, P. and Mertig, I. and Gy\"orffy, B. L.},
  journal = {Phys. Rev. B},
  volume = {84},
  issue = {7},
  pages = {075113},
  numpages = {12},
  year = {2011},
  month = {Aug},
  publisher = {American Physical Society},
  doi = {10.1103/PhysRevB.84.075113},
  url = {https://link.aps.org/doi/10.1103/PhysRevB.84.075113}
}

@article{Mitscherling2025,
  title = {Gauge-invariant projector calculus for quantum state geometry and applications to observables in crystals},
  author = {Mitscherling, Johannes and Avdoshkin, Alexander and Moore, Joel E.},
  journal = {Phys. Rev. B},
  volume = {112},
  issue = {8},
  pages = {085104},
  numpages = {16},
  year = {2025},
  month = {Aug},
  publisher = {American Physical Society},
  doi = {10.1103/qscv-qxqt},
  url = {https://link.aps.org/doi/10.1103/qscv-qxqt}
}

@article{Jarzynski2013,
  title = {Generating shortcuts to adiabaticity in quantum and classical dynamics},
  author = {Jarzynski, Christopher},
  journal = {Phys. Rev. A},
  volume = {88},
  issue = {4},
  pages = {040101},
  numpages = {5},
  year = {2013},
  month = {Oct},
  publisher = {American Physical Society},
  doi = {10.1103/PhysRevA.88.040101},
  url = {https://link.aps.org/doi/10.1103/PhysRevA.88.040101}
}

@article{Kato1950,
author = {Kato ,Tosio},
title = {{On the Adiabatic Theorem of Quantum Mechanics}},
journal = {Journal of the Physical Society of Japan},
volume = {5},
number = {6},
pages = {435-439},
year = {1950},
doi = {10.1143/JPSJ.5.435},
URL = { https://doi.org/10.1143/JPSJ.5.435},
eprint = { https://doi.org/10.1143/JPSJ.5.435}
}

@Article{Bradlyn2022,
	title={{Lecture notes on Berry phases and topology}},
	author={Barry Bradlyn and Mikel Iraola},
	journal={SciPost Phys. Lect. Notes},
	pages={51},
	year={2022},
	publisher={SciPost},
	doi={10.21468/SciPostPhysLectNotes.51},
	url={https://scipost.org/10.21468/SciPostPhysLectNotes.51},
}

@Article{Avron1987,
	title={{Adiabatic Theorems and Applications to the Quantum Hall Effect}},
	author={J. E. Avron and R. Seiler and L. G. Yaffe},
	journal={Commun. Math. Phys.},
	pages={33-49},
    volume={110},
	year={1987},
	publisher={Springer-Verlag},
	doi={10.1007/BF01209015},
	url={https://doi.org/10.1007/BF01209015},
}

@article{Nenciu1980,
doi = {10.1088/0305-4470/13/2/002},
url = {https://doi.org/10.1088/0305-4470/13/2/002},
year = {1980},
month = {feb},
publisher = {},
volume = {13},
number = {2},
pages = {L15},
author = {G. Nenciu},
title = {On the adiabatic theorem of quantum mechanics},
journal = {Journal of Physics A: Mathematical and General}
}

@article{Berry1984,
  title = {Quantal phase factors accompanying adiabatic changes},
  author = {Berry, Michael V.},
  journal = {Proc. R. Soc. Lond. A},
  volume = {392},
  pages = {45-57},
  year = {1984},
  doi = {10.1098/rspa.1984.0023},
  url = {http://doi.org/10.1098/rspa.1984.0023}
}

@article{Berry2009,
doi = {10.1088/1751-8113/42/36/365303},
url = {https://doi.org/10.1088/1751-8113/42/36/365303},
year = {2009},
month = {aug},
publisher = {},
volume = {42},
number = {36},
pages = {365303},
author = {Berry, M V},
title = {Transitionless quantum driving},
journal = {Journal of Physics A: Mathematical and Theoretical}
}

@article{Ahn2022,
  title={Riemannian geometry of resonant optical responses},
  author={Ahn, Junyeong and Guo, Guang-Yu and Nagaosa, Naoto and Vishwanath, Ashvin},
  journal={Nature Physics},
  volume={18},
  number={3},
  pages={290--295},
  year={2022},
  publisher={Nature Publishing Group}
}

@article{Avdoshkin2023,
  title = {Extrinsic geometry of quantum states},
  author = {Avdoshkin, Alexander and Popov, Fedor K.},
  journal = {Phys. Rev. B},
  volume = {107},
  issue = {24},
  pages = {245136},
  numpages = {17},
  year = {2023},
  month = {Jun},
  publisher = {American Physical Society},
  doi = {10.1103/PhysRevB.107.245136},
  url = {https://link.aps.org/doi/10.1103/PhysRevB.107.245136}
}

@article{Souza2008,
  title = {{Dichroic $f$-sum rule and the orbital magnetization of crystals}},
  author = {Souza, Ivo and Vanderbilt, David},
  journal = {Phys. Rev. B},
  volume = {77},
  issue = {5},
  pages = {054438},
  numpages = {13},
  year = {2008},
  month = {Feb},
  publisher = {American Physical Society},
  doi = {10.1103/PhysRevB.77.054438},
  url = {https://link.aps.org/doi/10.1103/PhysRevB.77.054438}
}

@article{Ceresoli2006,
  title = {{Orbital magnetization in crystalline solids: Multi-band insulators, Chern insulators, and metals}},
  author = {Ceresoli, Davide and Thonhauser, T. and Vanderbilt, David and Resta, R.},
  journal = {Phys. Rev. B},
  volume = {74},
  issue = {2},
  pages = {024408},
  numpages = {13},
  year = {2006},
  month = {Jul},
  publisher = {American Physical Society},
  doi = {10.1103/PhysRevB.74.024408},
  url = {https://link.aps.org/doi/10.1103/PhysRevB.74.024408}
}

@article{Resta2010,
doi = {10.1088/0953-8984/22/12/123201},
url = {https://doi.org/10.1088/0953-8984/22/12/123201},
year = {2010},
month = {mar},
publisher = {},
volume = {22},
number = {12},
pages = {123201},
author = {Resta, Raffaele},
title = {Electrical polarization and orbital magnetization: the modern theories},
journal = {Journal of Physics: Condensed Matter}
}

@article{Jiang2025,
doi = {10.1088/1361-6633/ade454},
url = {https://doi.org/10.1088/1361-6633/ade454},
year = {2025},
month = {jul},
publisher = {IOP Publishing},
volume = {88},
number = {7},
pages = {076502},
author = {Jiang, Yiyang and Holder, Tobias and Yan, Binghai},
title = {{Revealing quantum geometry in nonlinear quantum materials}},
journal = {Reports on Progress in Physics}
}

@misc{Yu2025,
      title={{Quantum Geometry in Quantum Materials}}, 
      author={Jiabin Yu and B. Andrei Bernevig and Raquel Queiroz and Enrico Rossi and P\"aivi T\"orm\"a and Bohm-Jung Yang},
      year={2025},
      eprint={2501.00098},
      archivePrefix={arXiv},
      primaryClass={cond-mat.mes-hall},
      url={https://arxiv.org/abs/2501.00098}, 
}

@misc{Gao2025,
      title={{Quantum Geometry Phenomena in Condensed Matter Systems}}, 
      author={Anyuan Gao and Naoto Nagaosa and Ni Ni and Su-Yang Xu},
      year={2025},
      eprint={2508.00469},
      archivePrefix={arXiv},
      primaryClass={cond-mat.str-el},
      url={https://arxiv.org/abs/2508.00469}, 
}

@misc{Verma2025,
      title={{Quantum Geometry: Revisiting electronic scales in quantum matter}}, 
      author={Nishchhal Verma and Philip J. W. Moll and Tobias Holder and Raquel Queiroz},
      year={2025},
      eprint={2504.07173},
      archivePrefix={arXiv},
      primaryClass={cond-mat.mtrl-sci},
      url={https://arxiv.org/abs/2504.07173}, 
}

@article{Liu2024,
    author = {Liu, Tianyu and Qiang, Xiao-Bin and Lu, Hai-Zhou and Xie, X C},
    title = {Quantum geometry in condensed matter},
    journal = {National Science Review},
    volume = {12},
    number = {3},
    pages = {nwae334},
    year = {2024},
    month = {09},
    issn = {2095-5138},
    doi = {10.1093/nsr/nwae334},
    url = {https://doi.org/10.1093/nsr/nwae334},
    eprint = {https://academic.oup.com/nsr/article-pdf/12/3/nwae334/59204701/nwae334.pdf},
}

@article{Xiao2007,
  title = {{Valley-Contrasting Physics in Graphene: Magnetic Moment and Topological Transport}},
  author = {Xiao, Di and Yao, Wang and Niu, Qian},
  journal = {Phys. Rev. Lett.},
  volume = {99},
  issue = {23},
  pages = {236809},
  numpages = {4},
  year = {2007},
  month = {Dec},
  publisher = {American Physical Society},
  doi = {10.1103/PhysRevLett.99.236809},
  url = {https://link.aps.org/doi/10.1103/PhysRevLett.99.236809}
}

@article{Avdoshkin2025,
  title = {{Multistate Geometry of Shift Current and Polarization}},
  author = {Avdoshkin, Alexander and Mitscherling, Johannes and Moore, Joel E.},
  journal = {Phys. Rev. Lett.},
  volume = {135},
  issue = {6},
  pages = {066901},
  numpages = {8},
  year = {2025},
  month = {Aug},
  publisher = {American Physical Society},
  doi = {10.1103/w761-8nf7},
  url = {https://link.aps.org/doi/10.1103/w761-8nf7}
}

@misc{Resta2017,
      title={{Geometrical meaning of the Drude weight and its relationship to orbital magnetization}}, 
      author={Raffaele Resta},
      year={2017},
      eprint={1703.00712},
      archivePrefix={arXiv},
      primaryClass={cond-mat.mtrl-sci},
      url={https://arxiv.org/abs/1703.00712}, 
}

@article{Resta2020,
  title = {Magnetic circular dichroism versus orbital magnetization},
  author = {Resta, Raffaele},
  journal = {Phys. Rev. Res.},
  volume = {2},
  issue = {2},
  pages = {023139},
  numpages = {6},
  year = {2020},
  month = {May},
  publisher = {American Physical Society},
  doi = {10.1103/PhysRevResearch.2.023139},
  url = {https://link.aps.org/doi/10.1103/PhysRevResearch.2.023139}
}

@article{Marzari1997,
  title = {Maximally localized generalized Wannier functions for composite energy bands},
  author = {Marzari, Nicola and Vanderbilt, David},
  journal = {Phys. Rev. B},
  volume = {56},
  issue = {20},
  pages = {12847--12865},
  numpages = {0},
  year = {1997},
  month = {Nov},
  publisher = {American Physical Society},
  doi = {10.1103/PhysRevB.56.12847},
  url = {https://link.aps.org/doi/10.1103/PhysRevB.56.12847}
}

@article{Lopez2012,
  title = {{Wannier-based calculation of the orbital magnetization in crystals}},
  author = {Lopez, M. G. and Vanderbilt, David and Thonhauser, T. and Souza, Ivo},
  journal = {Phys. Rev. B},
  volume = {85},
  issue = {1},
  pages = {014435},
  numpages = {12},
  year = {2012},
  month = {Jan},
  publisher = {American Physical Society},
  doi = {10.1103/PhysRevB.85.014435},
  url = {https://link.aps.org/doi/10.1103/PhysRevB.85.014435}
}

@article{Hanke2016,
  title = {{Role of Berry phase theory for describing orbital magnetism: From magnetic heterostructures to topological orbital ferromagnets}},
  author = {Hanke, J.-P. and Freimuth, F. and Nandy, A. K. and Zhang, H. and Bl\"ugel, S. and Mokrousov, Y.},
  journal = {Phys. Rev. B},
  volume = {94},
  issue = {12},
  pages = {121114},
  numpages = {5},
  year = {2016},
  month = {Sep},
  publisher = {American Physical Society},
  doi = {10.1103/PhysRevB.94.121114},
  url = {https://link.aps.org/doi/10.1103/PhysRevB.94.121114}
}

@article{Bianco2013,
  title = {{Orbital Magnetization as a Local Property}},
  author = {Bianco, Raffaello and Resta, Raffaele},
  journal = {Phys. Rev. Lett.},
  volume = {110},
  issue = {8},
  pages = {087202},
  numpages = {4},
  year = {2013},
  month = {Feb},
  publisher = {American Physical Society},
  doi = {10.1103/PhysRevLett.110.087202},
  url = {https://link.aps.org/doi/10.1103/PhysRevLett.110.087202}
}

@article{Nourafkan2014,
  title = {{Orbital magnetization of correlated electrons with arbitrary band topology}},
  author = {Nourafkan, R. and Kotliar, G. and Tremblay, A.-M. S.},
  journal = {Phys. Rev. B},
  volume = {90},
  issue = {12},
  pages = {125132},
  numpages = {10},
  year = {2014},
  month = {Sep},
  publisher = {American Physical Society},
  doi = {10.1103/PhysRevB.90.125132},
  url = {https://link.aps.org/doi/10.1103/PhysRevB.90.125132}
}

@article{Antebi2024,
  title = {{Drude weight of an interacting flat-band metal}},
  author = {Antebi, Ohad and Mitscherling, Johannes and Holder, Tobias},
  journal = {Phys. Rev. B},
  volume = {110},
  issue = {24},
  pages = {L241111},
  numpages = {7},
  year = {2024},
  month = {Dec},
  publisher = {American Physical Society},
  doi = {10.1103/PhysRevB.110.L241111},
  url = {https://link.aps.org/doi/10.1103/PhysRevB.110.L241111}
}

@article{Shindou2001,
  title = {{Orbital Ferromagnetism and Anomalous Hall Effect in Antiferromagnets on the Distorted fcc Lattice}},
  author = {Shindou, Ryuichi and Nagaosa, Naoto},
  journal = {Phys. Rev. Lett.},
  volume = {87},
  issue = {11},
  pages = {116801},
  numpages = {4},
  year = {2001},
  month = {Aug},
  publisher = {American Physical Society},
  doi = {10.1103/PhysRevLett.87.116801},
  url = {https://link.aps.org/doi/10.1103/PhysRevLett.87.116801}
}

@article{Feng2020,
  title={Topological magneto-optical effects and their quantization in noncoplanar antiferromagnets},
  author={Feng, Wanxiang and Hanke, Jan-Philipp and Zhou, Xiadong and Guo, Guang-Yu and Bl\"ugel, Stefan and Mokrousov, Yuriy and Yao, Yugui},
  journal={Nat. Commun.},
  volume={11},
  pages={118},
  year={2020},
  publisher={Nature Publishing Group},
  doi = {10.1038/s41467-019-13968-8},
  url = {https://doi.org/10.1038/s41467-019-13968-8}
}

@article{Zhou2016,
  title = {{Predicted Quantum Topological Hall Effect and Noncoplanar Antiferromagnetism in ${\mathrm{K}}_{0.5}{\mathrm{RhO}}_{2}$}},
  author = {Zhou, Jian and Liang, Qi-Feng and Weng, Hongming and Chen, Y. B. and Yao, Shu-Hua and Chen, Yan-Feng and Dong, Jinming and Guo, Guang-Yu},
  journal = {Phys. Rev. Lett.},
  volume = {116},
  issue = {25},
  pages = {256601},
  numpages = {5},
  year = {2016},
  month = {Jun},
  publisher = {American Physical Society},
  doi = {10.1103/PhysRevLett.116.256601},
  url = {https://link.aps.org/doi/10.1103/PhysRevLett.116.256601}
}

@article{Etxebarria2025,
author = "Etxebarria, Jesus and Perez-Mato, J. Manuel and Tasci, Emre S. and Elcoro, Luis",
title = "{Crystal tensor properties of magnetic materials with and without spin{--}orbit coupling. Application of spin point groups as approximate symmetries}",
journal = "Acta Crystallographica Section A",
year = "2025",
volume = "81",
number = "4",
pages = "317--338",
month = "Jul",
doi = {10.1107/S2053273325004127},
url = {https://doi.org/10.1107/S2053273325004127}
}

@article{Takagi2023,
  title={{Spontaneous topological Hall effect induced by non-coplanar antiferromagnetic order in intercalated van der Waals materials}},
  author={H. Takagi and R. Takagi and S. Minami and T. Nomoto and K. Oshishi and M.-T. Suzuki and Y. Yanagi and M. Hirayama and N. D. Khanh and K. Karube and H. Saito and D. Hashizume and R. Kiyanagi and Y. Tokura and R. Arita and T. Nakajima and S. Seki},
  journal={Nat. Phys.},
  volume={19},
  pages={967---968},
  year={2023},
  publisher={Nature Publishing Group},
  doi = {10.1038/s41567-023-02017-3},
  url = {https://doi.org/10.1038/s41567-023-02017-3}
}

@article{Khanh2025,
  title={{Spontaneous topological Hall effect induced by non-coplanar antiferromagnetic order in intercalated van der Waals materials}},
  author={Khanh, N. D. and Minami, S. and Hirschmann, M. M. and Nomoto, T. and Jiang, M.-C. and Yamada, R. and Heinsdorf, N. and Yamaguchi, D. and Hayashi, Y. and Okamura, Y. and Watanabe, H. and Guo, G.-Y. and Takahashi, Y. and Seki, S. and Taguchi, Y. and Tokura, Y. and Arita, R. and Hirschberger, M.},
  journal={Nat. Commun.},
  volume={16},
  pages={2654},
  year={2024},
  publisher={Nature Publishing Group},
  doi = {10.1038/s41467-025-57320-9},
  url = {https://doi.org/10.1038/s41467-025-57320-9}
}

@article{Hirschmann2024,
  title = {{Symmetry-enforced double Weyl points, multiband quantum geometry, and singular flat bands of doping-induced states at the Fermi level}},
  author = {Hirschmann, Moritz M. and Mitscherling, Johannes},
  journal = {Phys. Rev. Mater.},
  volume = {8},
  issue = {1},
  pages = {014201},
  numpages = {13},
  year = {2024},
  month = {Jan},
  publisher = {American Physical Society},
  doi = {10.1103/PhysRevMaterials.8.014201},
  url = {https://link.aps.org/doi/10.1103/PhysRevMaterials.8.014201}
}

@article{Resta2018,
doi = {10.1088/1361-648X/aade19},
url = {https://doi.org/10.1088/1361-648X/aade19},
year = {2018},
month = {sep},
publisher = {IOP Publishing},
volume = {30},
number = {41},
pages = {414001},
author = {Resta, Raffaele},
title = {Drude weight and superconducting weight},
journal = {Journal of Physics: Condensed Matter}
}

@article{Shinada2025,
  title = {{Quantum geometric bounds for observables: Linear responses, Drude weight, and orbital magnetization}},
  author = {Shinada, Koki and Nagaosa, Naoto},
  journal = {Phys. Rev. B},
  volume = {112},
  issue = {15},
  pages = {155158},
  numpages = {12},
  year = {2025},
  month = {Oct},
  publisher = {American Physical Society},
  doi = {10.1103/qxbl-qd4f},
  url = {https://link.aps.org/doi/10.1103/qxbl-qd4f}
}

@article{Pozo2020,
  title = {{Computing observables without eigenstates: Applications to Bloch Hamiltonians}},
  author = {Pozo, Oscar and de Juan, Fernando},
  journal = {Phys. Rev. B},
  volume = {102},
  issue = {11},
  pages = {115138},
  numpages = {12},
  year = {2020},
  month = {Sep},
  publisher = {American Physical Society},
  doi = {10.1103/PhysRevB.102.115138},
  url = {https://link.aps.org/doi/10.1103/PhysRevB.102.115138}
}

@article{Mera2022,
  title = {Nontrivial quantum geometry of degenerate flat bands},
  author = {Mera, Bruno and Mitscherling, Johannes},
  journal = {Phys. Rev. B},
  volume = {106},
  issue = {16},
  pages = {165133},
  numpages = {8},
  year = {2022},
  month = {Oct},
  publisher = {American Physical Society},
  doi = {10.1103/PhysRevB.106.165133},
  url = {https://link.aps.org/doi/10.1103/PhysRevB.106.165133}
}

@misc{Avdoshkin2024a,
      title={{Geometry of degenerate quantum states, configurations of $m$-planes and invariants on complex Grassmannians}}, 
      author={Alexander Avdoshkin},
      year={2024},
      eprint={2404.03234},
      archivePrefix={arXiv},
      primaryClass={quant-ph},
      url={https://arxiv.org/abs/2404.03234}, 
}

@misc{Ulrich2025,
      title={{Quantum Geometric Origin of the Intrinsic Nonlinear Hall Effect}}, 
      author={Yannis Ulrich and Johannes Mitscherling and Laura Classen and Andreas P. Schnyder},
      year={2025},
      eprint={2506.17386},
      archivePrefix={arXiv},
      primaryClass={cond-mat.mes-hall},
      url={https://arxiv.org/abs/2506.17386}, 
}

@article{Thonhauser2011,
author = {Thonhauser, T.},
title = {Theory of orbital magnetization in solids},
journal = {International Journal of Modern Physics B},
volume = {25},
number = {11},
pages = {1429-1458},
year = {2011},
doi = {10.1142/S0217979211058912},
URL = {https://doi.org/10.1142/S0217979211058912},
eprint = {https://doi.org/10.1142/S0217979211058912}
}

@article{Mehraeen2025,
  title = {{Quantum Response Theory and Momentum-Space Gravity}},
  author = {Mehraeen, M.},
  journal = {Phys. Rev. Lett.},
  volume = {135},
  issue = {15},
  pages = {156302},
  numpages = {7},
  year = {2025},
  month = {Oct},
  publisher = {American Physical Society},
  doi = {10.1103/t6nt-qzws},
  url = {https://link.aps.org/doi/10.1103/t6nt-qzws}
}

@article{Jankowski2024,
  title = {{Non-Abelian Hopf-Euler insulators}},
  author = {Jankowski, Wojciech J. and Morris, Arthur S. and Davoyan, Zory and Bouhon, Adrien and \"Unal, F. Nur and Slager, Robert-Jan},
  journal = {Phys. Rev. B},
  volume = {110},
  issue = {7},
  pages = {075135},
  numpages = {24},
  year = {2024},
  month = {Aug},
  publisher = {American Physical Society},
  doi = {10.1103/PhysRevB.110.075135},
  url = {https://link.aps.org/doi/10.1103/PhysRevB.110.075135}
}

@article{Jankowski2024a,
  title = {{Quantized Integrated Shift Effect in Multigap Topological Phases}},
  author = {Jankowski, Wojciech J. and Slager, Robert-Jan},
  journal = {Phys. Rev. Lett.},
  volume = {133},
  issue = {18},
  pages = {186601},
  numpages = {8},
  year = {2024},
  month = {Oct},
  publisher = {American Physical Society},
  doi = {10.1103/PhysRevLett.133.186601},
  url = {https://link.aps.org/doi/10.1103/PhysRevLett.133.186601}
}

@article{Schindler2025,
  title = {{Geometric Floquet Theory}},
  author = {Schindler, Paul M. and Bukov, Marin},
  journal = {Phys. Rev. X},
  volume = {15},
  issue = {3},
  pages = {031037},
  numpages = {21},
  year = {2025},
  month = {Aug},
  publisher = {American Physical Society},
  doi = {10.1103/7l91-gw77},
  url = {https://link.aps.org/doi/10.1103/7l91-gw77}
}

@article{Watanabe2024,
  title = {Symmetry analysis with spin crystallographic groups: Disentangling effects free of spin-orbit coupling in emergent electromagnetism},
  author = {Watanabe, Hikaru and Shinohara, Kohei and Nomoto, Takuya and Togo, Atsushi and Arita, Ryotaro},
  journal = {Phys. Rev. B},
  volume = {109},
  issue = {9},
  pages = {094438},
  numpages = {25},
  year = {2024},
  month = {Mar},
  publisher = {American Physical Society},
  doi = {10.1103/PhysRevB.109.094438},
  url = {https://link.aps.org/doi/10.1103/PhysRevB.109.094438}
}

@article{Pengfei2022,
  title = {Spin-Group Symmetry in Magnetic Materials with Negligible Spin-Orbit Coupling},
  author = {Liu, Pengfei and Li, Jiayu and Han, Jingzhi and Wan, Xiangang and Liu, Qihang},
  journal = {Phys. Rev. X},
  volume = {12},
  issue = {2},
  pages = {021016},
  numpages = {19},
  year = {2022},
  month = {Apr},
  publisher = {American Physical Society},
  doi = {10.1103/PhysRevX.12.021016},
  url = {https://link.aps.org/doi/10.1103/PhysRevX.12.021016}
}

@Article{Smejkal2022b,
author={{\v{S}}mejkal, Libor
and MacDonald, Allan H.
and Sinova, Jairo
and Nakatsuji, Satoru
and Jungwirth, Tomas},
title={Anomalous Hall antiferromagnets},
journal={Nature Reviews Materials},
year={2022},
month={Jun},
day={01},
volume={7},
number={6},
pages={482-496},
issn={2058-8437},
doi={10.1038/s41578-022-00430-3},
url={https://doi.org/10.1038/s41578-022-00430-3}
}

@Article{Hanke2017,
author={Hanke, Jan-Philipp
and Freimuth, Frank
and Bl{\"u}gel, Stefan
and Mokrousov, Yuriy},
title={Prototypical topological orbital ferromagnet $\gamma$-FeMn},
journal={Scientific Reports},
year={2017},
month={Jan},
day={20},
volume={7},
number={1},
pages={41078},
issn={2045-2322},
doi={10.1038/srep41078},
url={https://doi.org/10.1038/srep41078}
}

@article{Xiao2010,
  title = {Berry phase effects on electronic properties},
  author = {Xiao, Di and Chang, Ming-Che and Niu, Qian},
  journal = {Rev. Mod. Phys.},
  volume = {82},
  issue = {3},
  pages = {1959--2007},
  numpages = {0},
  year = {2010},
  month = {Jul},
  publisher = {American Physical Society},
  doi = {10.1103/RevModPhys.82.1959},
  url = {https://link.aps.org/doi/10.1103/RevModPhys.82.1959}
}

@article{Nagaosa2010,
  title = {{Anomalous Hall effect}},
  author = {Nagaosa, Naoto and Sinova, Jairo and Onoda, Shigeki and MacDonald, A. H. and Ong, N. P.},
  journal = {Rev. Mod. Phys.},
  volume = {82},
  issue = {2},
  pages = {1539--1592},
  numpages = {0},
  year = {2010},
  month = {May},
  publisher = {American Physical Society},
  doi = {10.1103/RevModPhys.82.1539},
  url = {https://link.aps.org/doi/10.1103/RevModPhys.82.1539}
}

@misc{Verma2024,
      title={{Instantaneous Response and Quantum Geometry of Insulators}}, 
      author={Nishchhal Verma and Raquel Queiroz},
      year={2025},
      eprint={2403.07052},
      archivePrefix={arXiv},
      primaryClass={cond-mat.mes-hall},
      url={https://arxiv.org/abs/2403.07052}, 
}

@article{Liu2025,
  title = {{Theory of Generalized Landau Levels and Its Implications for Non-Abelian States}},
  author = {Liu, Zhao and Mera, Bruno and Fujimoto, Manato and Ozawa, Tomoki and Wang, Jie},
  journal = {Phys. Rev. X},
  volume = {15},
  issue = {3},
  pages = {031019},
  numpages = {39},
  year = {2025},
  month = {Jul},
  publisher = {American Physical Society},
  doi = {10.1103/1zg9-qbd6},
  url = {https://link.aps.org/doi/10.1103/1zg9-qbd6}
}

@misc{Chen2025a,
      title={{The Effect of the Non-Abelian Quantum Metric on Superfluidity}}, 
      author={Kai Chen and Bishnu Karki and Pavan Hosur},
      year={2025},
      eprint={2501.16965},
      archivePrefix={arXiv},
      primaryClass={cond-mat.supr-con},
      url={https://arxiv.org/abs/2501.16965}, 
}

@article{Yao2008,
  title = {Valley-dependent optoelectronics from inversion symmetry breaking},
  author = {Yao, Wang and Xiao, Di and Niu, Qian},
  journal = {Phys. Rev. B},
  volume = {77},
  issue = {23},
  pages = {235406},
  numpages = {7},
  year = {2008},
  month = {Jun},
  publisher = {American Physical Society},
  doi = {10.1103/PhysRevB.77.235406},
  url = {https://link.aps.org/doi/10.1103/PhysRevB.77.235406}
}

@article{Roig2025,
  title = {Quasisymmetry-Constrained Spin Ferromagnetism in Altermagnets},
  author = {Roig, Merc\`e and Yu, Yue and Ekman, Rune C. and Kreisel, Andreas and Andersen, Brian M. and Agterberg, Daniel F.},
  journal = {Phys. Rev. Lett.},
  volume = {135},
  issue = {1},
  pages = {016703},
  numpages = {8},
  year = {2025},
  month = {Jul},
  publisher = {American Physical Society},
  doi = {10.1103/839n-rckn},
  url = {https://link.aps.org/doi/10.1103/839n-rckn}
}

@article{Pizzi2020,
doi = {10.1088/1361-648X/ab51ff},
url = {https://doi.org/10.1088/1361-648X/ab51ff},
year = {2020},
month = {jan},
publisher = {IOP Publishing},
volume = {32},
number = {16},
pages = {165902},
author = {Pizzi, Giovanni and Vitale, Valerio and Arita, Ryotaro and Bl\"ugel, Stefan and Freimuth, Frank and G\'eranton, Guillaume and Gibertini, Marco and Gresch, Dominik and Johnson, Charles and Koretsune, Takashi and Ibanez-Azpiroz, Julen and Lee, Hyungjun and Lihm, Jae-Mo and Marchand, Daniel and Marrazzo, Antimo and Mokrousov, Yuriy and Mustafa, Jamal I and Nohara, Yoshiro and Nomura, Yusuke and Paulatto, Lorenzo and Ponc\'e, Samuel and Ponweiser, Thomas and Qiao, Junfeng and Th\"ole, Florian and Tsirkin, Stepan S and Wierzbowska, Malgorzata and Marzari, Nicola and Vanderbilt, David and Souza, Ivo and Mostofi, Arash A and Yates, Jonathan R},
title = {Wannier90 as a community code: new features and applications},
journal = {Journal of Physics: Condensed Matter}
}

@misc{Ye2025,
      title={{Dominant orbital magnetization in the prototypical altermagnet MnTe}}, 
      author={Chao Chen Ye and Karma Tenzin and Jagoda Slawinska and Carmine Autieri},
      year={2025},
      eprint={2505.08675},
      archivePrefix={arXiv},
      primaryClass={cond-mat.mtrl-sci},
      url={https://arxiv.org/abs/2505.08675}, 
}

@Article{Martins2025,
	title={{Precise quantum-geometric electronic properties from first principles}},
	author={José Luís Martins and Carlos Loia Reis and Ivo Souza},
	journal={SciPost Phys.},
	volume={19},
	pages={109},
	year={2025},
	publisher={SciPost},
	doi={10.21468/SciPostPhys.19.4.109},
	url={https://scipost.org/10.21468/SciPostPhys.19.4.109},
}

@article{Tazuke2025,
  title = {{Formulation of the orbital magnetic moment in multiorbital tight-binding models: Application to the inverse Faraday effect}},
  author = {Tazuke, Kosuke and Morimoto, Takahiro and Kitamura, Sota},
  journal = {Phys. Rev. B},
  volume = {112},
  issue = {15},
  pages = {155134},
  numpages = {17},
  year = {2025},
  month = {Oct},
  publisher = {American Physical Society},
  doi = {10.1103/gmnv-cwvr},
  url = {https://link.aps.org/doi/10.1103/gmnv-cwvr}
}

@misc{Cysne2025,
      title={{Description of the orbital Hall effect from orbital magnetic moments of Bloch states: the role of a new correction term in bilayer systems}}, 
      author={Tarik P. Cysne and Ivo Souza and Tatiana G. Rappoport},
      year={2025},
      eprint={2511.03901},
      archivePrefix={arXiv},
      primaryClass={cond-mat.mes-hall},
      url={https://arxiv.org/abs/2511.03901}, 
}

@article{Chen2024,
  title = {{Enumeration and Representation Theory of Spin Space Groups}},
  author = {Chen, Xiaobing and Ren, Jun and Zhu, Yanzhou and Yu, Yutong and Zhang, Ao and Liu, Pengfei and Li, Jiayu and Liu, Yuntian and Li, Caiheng and Liu, Qihang},
  journal = {Phys. Rev. X},
  volume = {14},
  issue = {3},
  pages = {031038},
  numpages = {33},
  year = {2024},
  month = {Aug},
  publisher = {American Physical Society},
  doi = {10.1103/PhysRevX.14.031038},
  url = {https://link.aps.org/doi/10.1103/PhysRevX.14.031038}
}

@article{Zhu2025,
  title={Magnetic geometry induced quantum geometry and nonlinear transports},
  author={Zhu, Haiyuan and Li, Jiayu and Chen, Xiaobing and Yu, Yutong and Liu, Qihang},
  journal={Nat. Commun.},
  volume={16},
  pages={4882},
  year={2025}
}

\end{document}